\documentclass[onefignum,onetabnum]{siamart190516}
\usepackage{lipsum}
\usepackage{amsfonts}
\usepackage{graphicx}
\usepackage{epstopdf}
\usepackage{algorithmic}
\ifpdf
  \DeclareGraphicsExtensions{.eps,.pdf,.png,.jpg}
\else
  \DeclareGraphicsExtensions{.eps}
\fi

\newsiamremark{remark}{Remark}
\newsiamremark{hypothesis}{Hypothesis}
\crefname{hypothesis}{Hypothesis}{Hypotheses}
\newsiamthm{claim}{Claim}

\headers{Tensor Software}{C. Psarras, L. Karlsson, J. Li and P. Bientinesi}

\title{Tensor Software\thanks{\funding{This work was funded by the Deutsche Forschungsgemeinschaft (German Research Foundation) through grant IRTG 2379.}}}

\author{Christos Psarras\thanks{Aachen Institute for Advanced Study in Computational Engineering Science, RWTH Aachen University, Aachen, Germany
        (\email{psarras@aices.rwth-aachen.de}).}
   \and Lars Karlsson\footnotemark[3]
   \and Jiajia Li\thanks{Pacific Northwest National Laboratory, Richland, Washington and College of William \& Mary, Williamsburg, Virginia (\email{Jiajia.Li@pnnl.gov}).}
   \and Paolo Bientinesi\thanks{Department of Computing Science, Ume{\aa} Universitet, Ume{\aa}, Sweden
    (\email{larsk@cs.umu.se, pauldj@cs.umu.se}).}}

\usepackage{amsopn}

\ifpdf
\hypersetup{
  pdftitle={The landscape of software for tensor computations},
  pdfauthor={C. Psarras, L. Karlsson, J. Li and P. Bientinesi}
}
\fi

\definecolor{rwthblue}   {RGB}{  0,  84, 159}
\definecolor{rwthblue75}{RGB}{ 64, 127, 183}
\definecolor{rwthblue50}{RGB}{142, 186, 229}
\definecolor{rwthblue25}{RGB}{199, 221, 242}
\definecolor{rwthblue10}{RGB}{232, 241, 250}

\definecolor{rwthblack}  {RGB}{  0,   0,   0}
\definecolor{rwthblack75}{RGB}{100, 101, 103}
\definecolor{rwthblack50}{RGB}{156, 158, 159}
\definecolor{rwthblack25}{RGB}{207, 209, 210}
\definecolor{rwthblack10}{RGB}{236, 237, 237}

\definecolor{rwthdarkgray}{named}{rwthblack75}
\definecolor{rwthgray}{named}{rwthblack50}
\definecolor{rwthlightgray}{named}{rwthblack25}
\definecolor{rwthverylightgray}{named}{rwthblack10}

\definecolor{rwthmagenta}  {RGB}{227,   0, 102}
\definecolor{rwthmagenta75}{RGB}{233,  96, 136}
\definecolor{rwthmagenta50}{RGB}{241, 158, 177}
\definecolor{rwthmagenta25}{RGB}{249, 210, 218}
\definecolor{rwthmagenta10}{RGB}{253, 238, 240}

\definecolor{rwthyellow}  {RGB}{255, 237,   0}
\definecolor{rwthyellow75}{RGB}{255, 240,  85}
\definecolor{rwthyellow50}{RGB}{255, 245, 155}
\definecolor{rwthyellow25}{RGB}{255, 250, 209}
\definecolor{rwthyellow10}{RGB}{255, 253, 238}

\definecolor{rwthpetrol}   {RGB}{  0,  97, 101}
\definecolor{rwthpetrol75}{RGB}{ 45, 127, 131}
\definecolor{rwthpetrol50}{RGB}{125, 164, 167}
\definecolor{rwthpetrol25}{RGB}{191, 208, 209}
\definecolor{rwthpetrol10}{RGB}{230, 236, 236}

\definecolor{rwthturquoie}   {RGB}{  0, 152, 161}
\definecolor{rwthturquoise75}{RGB}{  0, 177, 183}
\definecolor{rwthturquoise50}{RGB}{137, 204, 207}
\definecolor{rwthturquoise25}{RGB}{202, 231, 231}
\definecolor{rwthturquoise10}{RGB}{235, 246, 246}

\definecolor{rwthgreen}  {RGB}{ 87, 171,  39}
\definecolor{rwthgreen75}{RGB}{141, 192,  96}
\definecolor{rwthgreen50}{RGB}{184, 214, 152}
\definecolor{rwthgreen25}{RGB}{221, 235, 206}
\definecolor{rwthgreen10}{RGB}{242, 247, 236}

\definecolor{rwthlightgreen}   {RGB}{189, 205,   0}
\definecolor{rwthlightgreen75}{RGB}{208, 217,  92}
\definecolor{rwthlightgreen50}{RGB}{224, 230, 154}
\definecolor{rwthlightgreen25}{RGB}{240, 243, 208}
\definecolor{rwthlightgreen10}{RGB}{249, 250, 237}

\definecolor{rwthorange}  {RGB}{246, 168,   0}
\definecolor{rwthorange75}{RGB}{250, 190,  80}
\definecolor{rwthorange50}{RGB}{253, 212, 143}
\definecolor{rwthorange25}{RGB}{254, 234, 201}
\definecolor{rwthorange10}{RGB}{255, 247, 234}

\definecolor{rwthred}  {RGB}{204,   7,  30}
\definecolor{rwthred75}{RGB}{216,  92,  65}
\definecolor{rwthred50}{RGB}{230, 150, 121}
\definecolor{rwthred25}{RGB}{243, 205, 187}
\definecolor{rwthred10}{RGB}{250, 235, 227}

\definecolor{rwthbordeaured}   {RGB}{161,  16,  53}
\definecolor{rwthbordeauxred75}{RGB}{182,  82,  86}
\definecolor{rwthbordeauxred50}{RGB}{205, 139, 135}
\definecolor{rwthbordeauxred25}{RGB}{229, 197, 192}
\definecolor{rwthbordeauxred10}{RGB}{245, 232, 229}

\definecolor{rwthviolet}  {RGB}{ 97,  33,  88}
\definecolor{rwthviolet75}{RGB}{131,  78, 117}
\definecolor{rwthviolet50}{RGB}{168, 133, 158}
\definecolor{rwthviolet25}{RGB}{210, 192, 205}
\definecolor{rwthviolet10}{RGB}{237, 229, 234}

\definecolor{rwthpurple}  {RGB}{122, 111, 172}
\definecolor{rwthpurple75}{RGB}{155, 145, 193}
\definecolor{rwthpurple50}{RGB}{188, 181, 215}
\definecolor{rwthpurple25}{RGB}{222, 218, 235}
\definecolor{rwthpurple10}{RGB}{242, 240, 247}

\usepackage{booktabs} % For formal tables
\usepackage{longtable}
\usepackage{hyperref}
\usepackage{graphicx}
\usepackage{pdflscape}

\def\CC{{C\nolinebreak[4]\hspace{-.05em}\raisebox{.4ex}{\tiny\bf ++}}}
\newcommand{\yes}{\checkmark}
\newcommand{\no}{\footnotesize{$-$}}

\usepackage{xspace}
\usepackage{textcomp}

\usepackage{multirow}
\usepackage{lscape}

\newcolumntype{g}{>{\columncolor{gray!25}}c}

\begin{document}

\title{The landscape of software for tensor computations}
%\subtitle{Current state of linear algebra languages and libraries.}

\maketitle

% REQUIRED
\begin{abstract}
  Tensors (also commonly seen as multi-linear operators or as multi-dimensional arrays) are ubiquitous in scientific computing
  and in data science, and so are the software efforts for tensor operations. Particularly in recent years, we have
  observed an explosion in libraries, compilers, packages, and toolboxes; unfortunately these efforts are very much
  scattered among the different scientific domains, and inevitably suffer from replication, suboptimal
  implementations, and in many cases, limited visibility.
  % dispersion of resources.
  % The software landscape for computations with tensors (or multi-way arrays) is scattered.
  % This leads to replication of effort, suboptimal implementations, and dispersion of resources.
  As a first step towards countering these inefficiencies, here we survey and loosely classify software packages related
  to tensor computations.
  Our aim is to assemble a comprehensive and up-to-date snapshot of the tensor software landscape, with the intention of
  helping both users and developers.
  Aware of the difficulties inherent in any multi-discipline survey,
  we very much welcome the reader's help in amending and expanding our software list, which currently features 80 projects.
  % we start counting from 0
\end{abstract}

% REQUIRED
\begin{keywords}
    tensor software, multi dimensional arrays, contractions, decompositions
\end{keywords}

% REQUIRED
\begin{AMS}
 68N15 % Programming languages
 68N20 % Compilers and interpreters
\end{AMS}

\section{Introduction}\label{sec:introduction}

Similar to matrices, tensors arise in a multitude of disciplines in engineering and science---for instance, in
computational chemistry, computational physics, chemometrics, data science, signal 
processing, and machine learning~\cite{Kolda,Sidiropoulos,Common,Bro,Rabanser,Cichocki}---and naturally, significant
effort goes into the development of numerical software.
However, in sharp contrast to the software landscape for (dense) matrix computations,
which is nicely layered and organized, 
that of tensor computations is fragmented and largely unstructured. 
Indeed, the tensor counterparts to the universally-used libraries such as BLAS---collection of building blocks---and
LAPACK---collection of solvers---are still missing. 

%software for tensor computations still lacks
%organization and structure: While (dense) linear algebra software is nicely layered and organized around well
%established sets of building blocks (BLAS) and solvers (e.g., LAPACK), the same is not true for tensors.

% Similar to matrix computations, tensor computations arise in a multitude of disciplines in engineering and science, 
% including (but not limited to) computational chemistry, computational physics, chemometrics, data science, signal
% processing, and machine learning~\cite{Kolda,Sidiropoulos,Common,Bro,Rabanser,Cichocki}.

When surveying the landscape of libraries, packages, compilers, and toolboxes\footnote{For the sake of simplicity, in the rest of this manuscript we will refer to all of these types of
  software simply as ``packages''.} for tensor computations, a massive replication of effort becomes apparent, and  
the absence of building blocks libraries is certainly one of the main causes for this. 
Other reasons are to be associated to the fact that tensor software is mostly driven
by applications, and is therefore scattered among different communities and scientific outlets.
It could be argued that matrix computations are possibly even more widespread,
yet a community effort made it  possible to create
``collection'' libraries and standardize interfaces already in the 1970s.
The profound difference is that while the
language (and notation) for linear algebra (i.e., matrix computations)
is quite consistent across disparate disciplines, the same cannot be further from the truth
when it comes to  the language of multi-linear algebra and multi-dimensional arrays (i.e., tensor computations).
Even the most basic concepts, such as the number and the length of the ``axes'' of a
tensor, have entirely different (and often conflicting) names in different disciplines.

In short, the development of tensor software has often been carried out independently in different communities, and even
within the same community there has not been any real coordinated effort.  
Motivated by this observation, we set out to survey the software landscape for tensor computations, aiming to 
%
% Computations with tensors (or multi-way arrays) have become prevalent in a variety of fields, ranging from computational
% chemistry and biology, to data science, signal processing, and machine learning
% \cite{Kolda,Sidiropoulos,Common,Bro,Rabanser,Cichocki}.
% Naturally, this 
% This has created an increased interest in the scientific community for software, capable of catering to their computational needs; as a result, myriads of software packages, libraries, extensions etc.\footnote{We will refer to all of them as packages from now on, to avoid repetition.} have been appearing in the software landscape.
% While on the one hand this shows that the community is lively and active, on the other hand, this development process inevitably ends up in replication of effort, suboptimal implementations and dispersion of resources, similar to the development of linear algebra-related routines and libraries before the introduction of BLAS and LAPACK in the 90s.
%
%Our aim is to survey the software landscape in an effort to
a) create awareness among users of the existing packages, and
b) guide the development at large. 
We see this survey as an essential step towards finding common ground between different applications, and towards identifying possible divisions of concerns, with the ultimate goal of defining a set of fundamental computational building blocks.

This survey takes up from a variety of communities, which often do not code in the same programming language, and, perhaps more importantly, present notable differences in the symbols and nomenclature they use to describe tensor operations.
In this document, we do \emph{not} address those differences, \emph{nor} aim to rank packages qualitatively, in any way.
Instead, we are merely attempting to put this diverse and large set of software packages on the map, with a loose classification of the functionality they provide.

This document is very much a work in progress, and we plan to keep it up-to-date by uploading new versions with some regularity.
To this end, we welcome and encourage input, contributions and corrections, to help create a more complete and fair snapshot of the current tensor software landscape.
We invite readers to send us contributions via email, and would greatly appreciate consulting the questionnaire in Appendix \ref{app:questionnaire}.

%%% Local Variables:
%%% mode: latex
%%% TeX-master: "../Tensor_SW"
%%% End:

\section{Software list}\label{sec:table}

We present a list of packages that support some form of tensor computations. 
To be considered for inclusion, a package must offer functionality in at least one of the following categories.
\begin{itemize}
\item \textbf{Data Manipulation} (DatM): Any operation related to the layout or storage of tensors,
  % ranging from simple representation, to
  such as tensor transposition, reshaping, conversion between different storage formats, \dots
\item \textbf{Element-Wise Operations} (EWOps): Any kind of element-by-element operation such as addition/subtraction,
  and/or reductions such as norms, min, max, \dots
\item \textbf{Contractions} (Con): General contractions between two or more tensors. Currently the survey does not differentiate between binary, ternary, or hypercontrations.
\item \textbf{Specific Contractions} (SpecCon): Specific operations that qualify as specific contractions, e.g.,
  Tensor Times Vector (TTV), Tensor Times Matrix (TTM), Matricized Tensor Times Khatri-Rao Product (MTTKRP), \dots
\item \textbf{Decompositions} (Decomp): At least one tensor decomposition, including but not limited to the Canonical Polyadic Decomposition (CPD or CP, also known as PARAllel FACtors analysis, PARAFAC), the Tucker Decomposition, Tensor Train, and their variants.
\end{itemize}

The following are the key aspects of a package that we aim to focus on.
\begin{itemize}
\item \textbf{Language}: What language is it written in, and, in the case of compilers/transpilers, what language does
  it generate code in (denoted by a $\rightarrow$)?
\item \textbf{Tensor type}: What type of tensor does it operate on? (e.g., Dense (D), Sparse (S), BlockSparse (BS), symmetric,
  supersymmetric, \dots) 
\item \textbf{Target system}: What types of computing architecture does it target? (e.g., CPU
  (C), GPU (G), Distributed Memory (D), \dots). Note: For CPU, there is currently no distinction between single-threaded
  and multi-threaded implementations.
  % since it is difficult to draw that distinction, given that most packages, if they don't explicitly employ SM practices, link to BLAS/LAPACK, which implicitly make use of parallelism.
\item \textbf{Functionality}: Which of the categories mentioned above does it support?
\end{itemize}

In the following tables, packages are listed alphabetically. 
For each package, Table~\ref{tab:big-table} provides when available a hyperlink to the source code (click on the package
name), and a reference to a publication or website. Additionally, each package is listed with an ID number. 

Tables~\ref{tab:fucntionality-dm},~\ref{tab:fucntionality-con}, and~\ref{tab:fucntionality-decomp} group all packages
according to the categores DatM, Con, and Decomp, respectively. 

Finally, Table~\ref{tab:fucntionality-complete} attempts to gather the ``more complete'' packages, i.e., those that offer support
for at least four out of the five categories described above. 
%Table~\ref{tab:fucntionality-spec} contains packages that offer a generalized form of a specific operation (e.g. tensor transposition or tensor contractions) in the form of a library.
%Tables~\ref{tab:fucntionality-complete} and~\ref{tab:fucntionality-spec}
%To some extent, these packages can be considered as mature, and it is likely that new developers can leverage them to build their own applications.

% TODO: order, precision (S/D/SC/DC), domain 

\begin{landscape}
\setlength{\LTcapwidth}{0.9\paperwidth}
\small
\begin{longtable}{ c l  c  c  c  c  c  c  l  l }

    \toprule 
    \multirow{2}{*}{\textbf{ID}} & \multirow{2}{*}{\textbf{Package Name}} & \multicolumn{5}{c}{\textbf{Functionality}} & \multirow{2}{*}{\textbf{Tensor Type}} & \multirow{2}{*}{\textbf{Platform}} & \multirow{2}{*}{\textbf{Language}} \\ \cmidrule{3-7}

    & & \textbf{DatM} & \textbf{EWOps} & \textbf{SpecCon} & \textbf{Con} & \textbf{Decomp} & &  & \\
    \midrule 
    \endhead

    \hypertarget{det:0}{}\hypertarget{det:1}{0} & \href{https://github.com/LLNL/acrotensor}{Acrotensor} \cite{Acrotensor} & \no & \no & \yes & \yes & \no & D & C, G & {\CC}\\ 
\rowcolor{gray!15} 
\hypertarget{det:2}{1} & \href{https://github.com/hpcgarage/AdaTM}{AdaTM} \cite{AdaTM} & \no & \no & \yes & \no & \yes & S & C & C\\ 
\hypertarget{det:3}{2} & \href{https://github.com/boostorg/ublas}{Boost.uBlas.Tensor} \cite{Boost.uBlas.Tensor} & \yes & \yes & \yes & \yes & \no & D & C & {\CC}\\ 
\rowcolor{gray!15} 
\hypertarget{det:4}{3} & \href{https://github.com/ValeevGroup/BTAS}{BTAS} \cite{BTAS} & \yes & \yes & \yes & \yes & \yes & nan & C & {\CC}\\ 
\hypertarget{det:5}{4} & \href{https://github.com/kimjsung/CGO2019-AE}{COGENT} \cite{COGENT} & \no & \no & \yes & \yes & \no & D & G & Python {\textrightarrow} CUDA\\ 
\rowcolor{gray!15} 
\hypertarget{det:6}{5} & COMET \cite{COMET} & \no & \no & \yes & \yes & \no & S & C & {\CC} {\textrightarrow} {\CC}\\ 
\hypertarget{det:7}{6} & \href{https://github.com/jcmgray/cotengra}{CoTenGra} \cite{CoTenGra} & \no & \no & \yes & \yes & \no & D & C, D, G & Python  \\ 
\rowcolor{gray!15} 
\hypertarget{det:8}{7} & \href{https://github.com/HPAC/CP-CALS}{CP-CALS} \cite{CP-CALS} & \no & \no & \yes & \no & \yes & D & C, G & {\CC}, Mat\textsuperscript{i}\\ 
\hypertarget{det:9}{8} & \href{https://github.com/ZacBlanco/cstf}{CSTF} \cite{CSTF} & \no & \no & \no & \no & \yes & S & D & Scala\\ 
\rowcolor{gray!15} 
\hypertarget{det:10}{9} & \href{https://developer.nvidia.com/cutensor/downloads}{CuTensor} \cite{CuTensor} & \yes & \yes & \yes & \yes & \no & D & G & C, CUDA\\ 
\hypertarget{det:11}{10} & \href{https://github.com/ap-hynninen/cutt}{cuTT} \cite{cuTT} & \yes & \no & \no & \no & \no & D & G & {\CC}, CUDA\\ 
\rowcolor{gray!15} 
\hypertarget{det:12}{11} & \href{https://github.com/cyclops-community/ctf}{Cyclops} \cite{Cyclops} & \yes & \yes & \yes & \yes & \no & S, D & C, D, G & {\CC}\\ 
\hypertarget{det:13}{12} & \href{https://datalab.snu.ac.kr/dtucker/resources/DTucker-v1.0.tar.gz}{D-Tucker} \cite{D-Tucker} & \no & \no & \no & \no & \yes & D & C & Matlab\\ 
\rowcolor{gray!15} 
\hypertarget{det:14}{13} & \href{http://www.joonheechoi.com/research.}{DFacTo} \cite{DFacTo} & \no & \no & \no & \no & \yes & S & C, D & {\CC}\\ 
\hypertarget{det:15}{14} & \href{https://gitlab.com/libeigen/eigen/-/tree/master/unsupported/Eigen/CXX11/src/Tensor}{Eigen Tensor} \cite{EigenTensor} & \yes & \yes & \yes & \yes & \no & D & C, G & {\CC}\\ 
\rowcolor{gray!15} 
\hypertarget{det:16}{15} & \href{https://github.com/ORNL-QCI/exatn}{ExaTN} \cite{ExaTN} & \yes & \yes & \yes & \yes & \yes & D & C, D, G & {\CC}, Py\textsuperscript{i}\\ 
\hypertarget{det:17}{16} & \href{https://github.com/romeric/Fastor}{Fastor} \cite{Fastor} & \yes & \yes & \yes & \yes & \no & D & C & {\CC}\\ 
\rowcolor{gray!15} 
\hypertarget{det:18}{17} & \href{https://gitlab.com/wlandry/ftensor}{FTensor} \cite{FTensor} & \yes & \yes & \yes & \yes & \no & D & C & {\CC}\\ 
\hypertarget{det:19}{18} & \href{https://gitlab.com/tensors/genten}{Genten} \cite{Genten} & \no & \no & \no & \no & \yes & D, S & C, G & {\CC}\\ 
\rowcolor{gray!15} 
\hypertarget{det:20}{19} & GigaTensor \cite{GigaTensor} & \no & \no & \no & \no & \yes & S & C, D & Unknown\\ 
\hypertarget{det:21}{20} & \href{https://github.com/HPAC/hptt}{HPTT} \cite{HPTT} & \yes & \no & \no & \no & \no & D & C & {\CC}, Python\textsuperscript{i}, C\textsuperscript{i}\\ 
\rowcolor{gray!15} 
\hypertarget{det:22}{21} & \href{https://github.com/ITensor/ITensor}{ITensor} \cite{ITensor} & \no & \yes & \yes & \yes & \yes & D, BS & C, G\textsuperscript{x} & {\CC}, Julia\\ 
\hypertarget{det:23}{22} & \href{https://github.com/epifanovsky/libtensor}{libtensor} \cite{libtensor} & \no & \no & \yes & \yes & \no & D, BS & C & {\CC}\\ 
\rowcolor{gray!15} 
\hypertarget{det:24}{23} & \href{https://code.google.com/archive/p/ltensor/source/default/source}{Ltensor} \cite{Ltensor} & \no & \no & \yes & \yes & \no & D & C & {\CC}\\ 
\hypertarget{det:25}{24} & \href{https://nl.mathworks.com/downloads/}{MATLAB} \cite{MATLAB} & \yes & \yes & \yes & \yes & \no & D & C & Matlab\\ 
\rowcolor{gray!15} 
\hypertarget{det:26}{25} & \href{https://dl.bintray.com/boostorg/release/1.75.0/source/}{MultiArray} \cite{MultiArray} & \yes & \no & \no & \no & \no & D & C & {\CC}\\ 
\hypertarget{det:27}{26} & \href{https://cran.r-project.org/web/packages/multiway/index.html}{multiway} \cite{multiway} & \no & \no & \no & \no & \yes & D & C & R\\ 
\rowcolor{gray!15} 
\hypertarget{det:28}{27} & \href{http://www.models.life.ku.dk/nwaytoolbox/download}{N-way toolbox} \cite{N-waytoolbox} & \no & \no & \no & \no & \yes & D & C & Matlab\\ 
\hypertarget{det:29}{28} & \href{https://physics.paperswithcode.com/paper/ncon-a-tensor-network-contractor-for-matlab#code}{NCON} \cite{NCON} & \no & \no & \yes & \yes & \no & D & C & Matlab\\ 
\rowcolor{gray!15} 
\hypertarget{det:30}{29} & \href{https://github.com/TensorCon/netcon}{netcon} \cite{netcon} & \no & \no & \yes & \yes & \no & D & C & Matlab\\ 
\hypertarget{det:31}{30} & \href{https://github.com/numpy/numpy}{NumPy} \cite{NumPy} & \yes & \yes & \yes & \yes & \no & D & C & Python\\ 
\rowcolor{gray!15} 
\hypertarget{det:32}{31} & \href{https://github.com/IBM/ocean-tensor-package}{Ocean} \cite{Ocean} & \yes & \yes & \no & \no & \no & D & C, G & C, Py\textsuperscript{i}\\ 
\hypertarget{det:33}{32} & \href{https://www.cs.ucr.edu/~epapalex/src/parCube.zip}{ParCube} \cite{ParCube} & \no & \no & \no & \no & \yes & S & C & Matlab\\ 
\rowcolor{gray!15} 
\hypertarget{det:34}{33} & \href{https://github.com/neurocom/partensor-toolbox}{ParTensor} \cite{ParTensor} & \no & \no & \no & \no & \yes & D & C, G & {\CC}\\ 
\hypertarget{det:35}{34} & \href{https://github.com/hpcgarage/ParTI}{ParTI!} \cite{ParTI} & \yes & \yes & \yes & \no & \yes & S & C, G & C, CUDA, Mat\textsuperscript{x}\\ 
\rowcolor{gray!15} 
\hypertarget{det:36}{35} & \href{https://github.com/ramkikannan/planc}{PLANC} \cite{PLANC} & \no & \no & \no & \no & \yes & D & C, D & {\CC}\\ 
\hypertarget{det:37}{36} & \href{https://eigenvector.com/software/pls-toolbox/}{PLS toolbox} \cite{PLStoolbox} & \no & \no & \no & \no & \yes & D & C & Matlab\\ 
\rowcolor{gray!15} 
\hypertarget{det:38}{37} & \href{https://code.google.com/archive/p/pytensor/source/default/source}{Pytensor} \cite{Pytensor} & \yes & \yes & \yes & \yes & \yes & D, S & C & Python\\ 
\hypertarget{det:39}{38} & \href{https://github.com/pytorch/pytorch}{PyTorch} \cite{PyTorch} & \yes & \yes & \yes & \yes & \no & D, S & C, G & Python, {\CC}, CUDA\\ 
\rowcolor{gray!15} 
\hypertarget{det:40}{39} & \href{https://github.com/jcmgray/quimb}{quimb} \cite{quimb} & \no & \no & \yes & \yes & \no & D & C, D, G & Python\\ 
\hypertarget{det:41}{40} & \href{https://github.com/jamesyili/rTensor}{rTensor} \cite{rTensor} & \yes & \yes & \yes & \no & \yes & D & C & R\\ 
\rowcolor{gray!15} 
\hypertarget{det:42}{41} & \href{https://github.com/erichson/rTensor}{rTensor (randomized)} \cite{rTensor(randomized)} & \no & \no & \no & \no & \yes & D & C & Python\\ 
\hypertarget{det:43}{42} & \href{https://github.com/mnick/scikit-tensor}{scikit-tensor} \cite{scikit-tensor} & \yes & \yes & \yes & \no & \yes & D, S & C & Python\\ 
\rowcolor{gray!15} 
\hypertarget{det:44}{43} & \href{https://github.com/PGelss/scikit_tt}{Scikit-TT} \cite{Scikit-TT} & \no & \no & \no & \no & \yes & D & C & Python\\ 
\hypertarget{det:45}{44} & \href{https://github.com/dehuacheng/SpAls}{SPALS} \cite{SPALS} & \no & \no & \no & \no & \yes & S & C & {\CC}\\ 
\rowcolor{gray!15} 
\hypertarget{det:46}{45} & \href{https://github.com/kperros/SPARTan}{SPARTan} \cite{SPARTan} & \no & \no & \no & \no & \yes & S & C & Matlab\\ 
\hypertarget{det:47}{46} & \href{https://github.com/ShadenSmith/splatt}{SPLATT} \cite{SPLATT} & \no & \no & \yes & \no & \yes & S & C, D & C, {\CC}, Oct, Mat\\ 
\rowcolor{gray!15} 
\hypertarget{det:48}{47} & \href{https://susmost.com/downloads.html}{SuSMoST} \cite{SuSMoST} & \no & \no & \no & \no & \yes & D & C & Python\\ 
\hypertarget{det:49}{48} & \href{https://github.com/Bihaqo/t3f}{T3F} \cite{T3F} & \yes & \yes & \no & \no & \yes & D & C, G & Python\\ 
\rowcolor{gray!15} 
\hypertarget{det:50}{49} & \href{https://github.com/tensor-compiler/taco}{TACO} \cite{TACO} & \yes & \yes & \yes & \yes & \no & D, S & C, G & {\CC}, {\CC} {\textrightarrow} {\CC}\\ 
\hypertarget{det:51}{50} & \href{https://github.com/DmitryLyakh/TAL_SH}{TAL{\_}SH} \cite{TAL_SH} & \yes & \yes & \yes & \yes & \no & D & C, G & C, {\CC}, Fort\\ 
\rowcolor{gray!15} 
\hypertarget{det:52}{51} & \href{https://github.com/devinamatthews/tblis}{TBlis} \cite{TBlis} & \yes & \yes & \yes & \yes & \no & D & C & {\CC}\\ 
\hypertarget{det:53}{52} & \href{https://github.com/HPAC/tccg}{TCCG} \cite{TCCG} & \no & \no & \yes & \yes & \no & D
 & C & {\CC}\\ 
\rowcolor{gray!15} 
\hypertarget{det:54}{53} & \href{https://github.com/springer13/tcl}{TCL} \cite{TCL} & \no & \no & \yes & \yes & \no & D & C & {\CC}, Python\textsuperscript{i}\\ 
\hypertarget{det:55}{54} & \href{https://github.com/andrewssobral/TDALAB}{TDALAB} \cite{TDALAB} & \no & \no & \no & \no & \yes & D, S & C & Matlab, GUI\\ 
\rowcolor{gray!15} 
\hypertarget{det:56}{55} & \href{https://github.com/tenpy/tenpy}{TeNPy} \cite{TeNPy} & \no & \no & \no & \no & \yes & D & C & Python\\ 
\hypertarget{det:57}{56} & \href{https://github.com/felipebottega/Tensor-Fox}{Tensor Fox} \cite{TensorFox} & \no & \no & \no & \no & \yes & D, S & C & Python, Matlab\\ 
\rowcolor{gray!15} 
\hypertarget{det:58}{57} & \href{http://www.gipsa-lab.fr/~pierre.comon/TensorPackage/tensorPackage.html}{Tensor package} \cite{Tensorpackage} & \no & \no & \no & \no & \yes & D & C & Matlab\\ 
\hypertarget{det:59}{58} & \href{https://gitlab.com/tensors/tensor_toolbox}{Tensor Toolbox} \cite{TensorToolbox} & \yes & \yes & \yes & \yes & \yes & D, S & C & Matlab\\ 
\rowcolor{gray!15} 
\hypertarget{det:60}{59} & \href{https://github.com/cyclops-community/tensor_decomposition}{tensor{\_}decomposition} \cite{tensor_decomposition} & \no & \no & \no & \no & \yes & D & C, D & Python\\ 
\hypertarget{det:61}{60} & \href{https://github.com/phananhhuy/TensorBox}{TensorBox} \cite{TensorBox} & \no & \no & \no & \no & \yes & D, S & C & Matlab\\ 
\rowcolor{gray!15} 
\hypertarget{det:62}{61} & \href{https://github.com/Large-Scale-Tensor-Decomposition/tensorD}{TensorD} \cite{TensorD} & \no & \no & \no & \no & \yes & D & C, G & Python\\ 
\hypertarget{det:63}{62} & \href{https://github.com/tensorflow/tensorflow}{TensorFlow} \cite{TensorFlow} & \yes & \yes & \yes & \yes & \no & D, S & C, D, G & {\CC}, Python\\ 
\rowcolor{gray!15} 
\hypertarget{det:64}{63} & \href{https://www.tensorlab.net/}{TensorLab} \cite{TensorLab} & \no & \no & \no & \no & \yes & D, S & C & Matlab\\ 
\hypertarget{det:65}{64} & \href{https://www.tensorlabplus.net/}{TensorLab+} \cite{TensorLab+} & \no & \no & \no & \no & \yes & D, S & C & Matlab\\ 
\rowcolor{gray!15} 
\hypertarget{det:66}{65} & \href{https://github.com/tensorly/tensorly}{TensorLy} \cite{TensorLy} & \yes & \yes & \yes & \yes & \yes & D & C, G & Python\\ 
\hypertarget{det:67}{66} & \href{https://github.com/google/TensorNetwork}{TensorNetwork} \cite{TensorNetwork} & \no & \no & \yes & \yes & \no & D, S & C, G & Python\\ 
\rowcolor{gray!15} 
\hypertarget{det:68}{67} & \href{https://github.com/Jutho/TensorOperations.jl}{TensorOperations.jl} \cite{TensorOperations.jl} & \yes & \yes & \yes & \yes & \no & D & C, G & Julia\\ 
\hypertarget{det:69}{68} & \href{https://www.tensortrace.com/downloads}{TensorTrace} \cite{TensorTrace} & \no & \no & \yes & \yes & \no & D & C & GUI {\textrightarrow} Py, Jul, Mat\\ 
\rowcolor{gray!15} 
\hypertarget{det:70}{69} & \href{https://github.com/cran/ThreeWay}{Three-Way} \cite{Three-Way} & \no & \no & \yes & \yes & \yes & D & C & R\\ 
\hypertarget{det:71}{70} & \href{https://github.com/ValeevGroup/tiledarray}{TiledArray} \cite{TiledArray} & \yes & \yes & \yes & \yes & \no & D, BS & C, D & {\CC}\\ 
\rowcolor{gray!15} 
\hypertarget{det:72}{71} & \href{https://github.com/andrewdarmawan/tncontract}{tncontract} \cite{tncontract} & \no & \no & \yes & \yes & \no & D & C & Python\\ 
\hypertarget{det:73}{72} & \href{https://github.com/ycyuustc/matlab}{TNR} \cite{TNR} & \no & \no & \no & \no & \yes & D & C & Matlab\\ 
\rowcolor{gray!15} 
\hypertarget{det:74}{73} & \href{https://github.com/jemisjoky/TorchMPS}{TorchMPS} \cite{TorchMPS} & \no & \no & \yes & \yes & \no & D & C & Python\\ 
\hypertarget{det:75}{74} & \href{https://github.com/oseledets/TT-Toolbox}{TT-Toolbox} \cite{TT-Toolbox} & \yes & \yes & \no & \no & \yes & D & C, D\textsuperscript{x}, G\textsuperscript{x} & Matlab, Python\\ 
\rowcolor{gray!15} 
\hypertarget{det:76}{75} & \href{https://github.com/HPAC/TTC}{TTC} \cite{TTC} & \yes & \no & \no & \no & \no & D & C & Python {\textrightarrow} {\CC}\\ 
\hypertarget{det:77}{76} & \href{https://github.com/bassoy/ttv}{TTV} \cite{TTV} & \no & \no & \yes & \no & \no & D & C & {\CC}\\ 
\rowcolor{gray!15} 
\hypertarget{det:78}{77} & \href{https://github.com/apache/tvm}{TVM} \cite{TVM} & \no & \yes & \no & \no & \no & D, S & C, G & Python\\ 
\hypertarget{det:79}{78} & \href{https://gitlab.com/uni10/uni10/}{Uni10} \cite{Uni10} & \yes & \yes & \yes & \yes & \no & D & C, G\textsuperscript{x} & {\CC}\\ 
\rowcolor{gray!15} 
\hypertarget{det:80}{79} & \href{https://git.hemio.de/xerus/xerus/}{xerus} \cite{xerus} & \no & \no & \yes & \yes & \yes & D, S & C,  & {\CC}\\

    \bottomrule
\caption{Main list. Packages are sorted alphabetically by name. We use an i superscript to denote an interface to the specific language. We use the x superscript to denote an experimental feature.}
\label{tab:big-table}
\end{longtable}
\end{landscape}

\medskip
\begin{longtable}{llccccc}
    \toprule
    ID & Name & \multicolumn{5}{c}{Functionality} \\
    & & DatM & EWOps & SpecCon & Con & Decomp \\
    \midrule
    \endhead
    \hyperlink{det:2}{2} & Boost.uBlas.Tensor & \yes & \yes & \yes & \yes & \no\\ 
\rowcolor{gray!15} 
\hyperlink{det:3}{3} & BTAS & \yes & \yes & \yes & \yes & \yes\\ 
\hyperlink{det:9}{9} & CuTensor & \yes & \yes & \yes & \yes & \no\\ 
\rowcolor{gray!15} 
\hyperlink{det:10}{10} & cuTT & \yes & \no & \no & \no & \no\\ 
\hyperlink{det:11}{11} & Cyclops & \yes & \yes & \yes & \yes & \no\\ 
\rowcolor{gray!15} 
\hyperlink{det:14}{14} & Eigen Tensor & \yes & \yes & \yes & \yes & \no\\ 
\hyperlink{det:15}{15} & ExaTN & \yes & \yes & \yes & \yes & \yes\\ 
\rowcolor{gray!15} 
\hyperlink{det:16}{16} & Fastor & \yes & \yes & \yes & \yes & \no\\ 
\hyperlink{det:17}{17} & FTensor & \yes & \yes & \yes & \yes & \no\\ 
\rowcolor{gray!15} 
\hyperlink{det:20}{20} & HPTT & \yes & \no & \no & \no & \no\\ 
\hyperlink{det:24}{24} & MATLAB & \yes & \yes & \yes & \yes & \no\\ 
\rowcolor{gray!15} 
\hyperlink{det:25}{25} & MultiArray & \yes & \no & \no & \no & \no\\ 
\hyperlink{det:30}{30} & NumPy & \yes & \yes & \yes & \yes & \no\\ 
\rowcolor{gray!15} 
\hyperlink{det:31}{31} & Ocean & \yes & \yes & \no & \no & \no\\ 
\hyperlink{det:34}{34} & ParTI! & \yes & \yes & \yes & \no & \yes\\ 
\rowcolor{gray!15} 
\hyperlink{det:37}{37} & Pytensor & \yes & \yes & \yes & \yes & \yes\\ 
\hyperlink{det:38}{38} & PyTorch & \yes & \yes & \yes & \yes & \no\\ 
\rowcolor{gray!15} 
\hyperlink{det:40}{40} & rTensor & \yes & \yes & \yes & \no & \yes\\ 
\hyperlink{det:42}{42} & scikit-tensor & \yes & \yes & \yes & \no & \yes\\ 
\rowcolor{gray!15} 
\hyperlink{det:48}{48} & T3F & \yes & \yes & \no & \no & \yes\\ 
\hyperlink{det:49}{49} & TACO & \yes & \yes & \yes & \yes & \no\\ 
\rowcolor{gray!15} 
\hyperlink{det:50}{50} & TAL{\_}SH & \yes & \yes & \yes & \yes & \no\\ 
\hyperlink{det:51}{51} & TBlis & \yes & \yes & \yes & \yes & \no\\ 
\rowcolor{gray!15} 
\hyperlink{det:58}{58} & Tensor Toolbox & \yes & \yes & \yes & \yes & \yes\\ 
\hyperlink{det:62}{62} & TensorFlow & \yes & \yes & \yes & \yes & \no\\ 
\rowcolor{gray!15} 
\hyperlink{det:65}{65} & TensorLy & \yes & \yes & \yes & \yes & \yes\\ 
\hyperlink{det:67}{67} & TensorOperations.jl & \yes & \yes & \yes & \yes & \no\\ 
\rowcolor{gray!15} 
\hyperlink{det:70}{70} & TiledArray & \yes & \yes & \yes & \yes & \no\\ 
\hyperlink{det:74}{74} & TT-Toolbox & \yes & \yes & \no & \no & \yes\\ 
\rowcolor{gray!15} 
\hyperlink{det:75}{75} & TTC & \yes & \no & \no & \no & \no\\ 
\hyperlink{det:78}{78} & Uni10 & \yes & \yes & \yes & \yes & \no\\ 

    \bottomrule
    \caption{\label{tab:fucntionality-dm}Packages that support Data Manipulation (DatM).}
\end{longtable}
\newpage
\begin{longtable}{llccccc}
    \toprule
    ID & Name & \multicolumn{5}{c}{Functionality} \\
    & & DatM & EWOps & SpecCon & Con & Decomp \\
    \midrule
    \endhead
    \hyperlink{det:0}{0} & Acrotensor & \no & \no & \yes & \yes & \no\\ 
\rowcolor{gray!15} 
\hyperlink{det:2}{2} & Boost.uBlas.Tensor & \yes & \yes & \yes & \yes & \no\\ 
\hyperlink{det:3}{3} & BTAS & \yes & \yes & \yes & \yes & \yes\\ 
\rowcolor{gray!15} 
\hyperlink{det:4}{4} & COGENT & \no & \no & \yes & \yes & \no\\ 
\hyperlink{det:5}{5} & COMET & \no & \no & \yes & \yes & \no\\ 
\rowcolor{gray!15} 
\hyperlink{det:6}{6} & CoTenGra & \no & \no & \yes & \yes & \no\\ 
\hyperlink{det:9}{9} & CuTensor & \yes & \yes & \yes & \yes & \no\\ 
\rowcolor{gray!15} 
\hyperlink{det:11}{11} & Cyclops & \yes & \yes & \yes & \yes & \no\\ 
\hyperlink{det:14}{14} & Eigen Tensor & \yes & \yes & \yes & \yes & \no\\ 
\rowcolor{gray!15} 
\hyperlink{det:15}{15} & ExaTN & \yes & \yes & \yes & \yes & \yes\\ 
\hyperlink{det:16}{16} & Fastor & \yes & \yes & \yes & \yes & \no\\ 
\rowcolor{gray!15} 
\hyperlink{det:17}{17} & FTensor & \yes & \yes & \yes & \yes & \no\\ 
\hyperlink{det:21}{21} & ITensor & \no & \yes & \yes & \yes & \yes\\ 
\rowcolor{gray!15} 
\hyperlink{det:22}{22} & libtensor & \no & \no & \yes & \yes & \no\\ 
\hyperlink{det:23}{23} & Ltensor & \no & \no & \yes & \yes & \no\\ 
\rowcolor{gray!15} 
\hyperlink{det:24}{24} & MATLAB & \yes & \yes & \yes & \yes & \no\\ 
\hyperlink{det:28}{28} & NCON & \no & \no & \yes & \yes & \no\\ 
\rowcolor{gray!15} 
\hyperlink{det:29}{29} & netcon & \no & \no & \yes & \yes & \no\\ 
\hyperlink{det:30}{30} & NumPy & \yes & \yes & \yes & \yes & \no\\ 
\rowcolor{gray!15} 
\hyperlink{det:37}{37} & Pytensor & \yes & \yes & \yes & \yes & \yes\\ 
\hyperlink{det:38}{38} & PyTorch & \yes & \yes & \yes & \yes & \no\\ 
\rowcolor{gray!15} 
\hyperlink{det:39}{39} & quimb & \no & \no & \yes & \yes & \no\\ 
\hyperlink{det:49}{49} & TACO & \yes & \yes & \yes & \yes & \no\\ 
\rowcolor{gray!15} 
\hyperlink{det:50}{50} & TAL{\_}SH & \yes & \yes & \yes & \yes & \no\\ 
\hyperlink{det:51}{51} & TBlis & \yes & \yes & \yes & \yes & \no\\ 
\rowcolor{gray!15} 
\hyperlink{det:52}{52} & TCCG & \no & \no & \yes & \yes & \no\\ 
\hyperlink{det:53}{53} & TCL & \no & \no & \yes & \yes & \no\\ 
\rowcolor{gray!15} 
\hyperlink{det:58}{58} & Tensor Toolbox & \yes & \yes & \yes & \yes & \yes\\ 
\hyperlink{det:62}{62} & TensorFlow & \yes & \yes & \yes & \yes & \no\\ 
\rowcolor{gray!15} 
\hyperlink{det:65}{65} & TensorLy & \yes & \yes & \yes & \yes & \yes\\ 
\hyperlink{det:66}{66} & TensorNetwork & \no & \no & \yes & \yes & \no\\ 
\rowcolor{gray!15} 
\hyperlink{det:67}{67} & TensorOperations.jl & \yes & \yes & \yes & \yes & \no\\ 
\hyperlink{det:68}{68} & TensorTrace & \no & \no & \yes & \yes & \no\\ 
\rowcolor{gray!15} 
\hyperlink{det:69}{69} & Three-Way & \no & \no & \yes & \yes & \yes\\ 
\hyperlink{det:70}{70} & TiledArray & \yes & \yes & \yes & \yes & \no\\ 
\rowcolor{gray!15} 
\hyperlink{det:71}{71} & tncontract & \no & \no & \yes & \yes & \no\\ 
\hyperlink{det:73}{73} & TorchMPS & \no & \no & \yes & \yes & \no\\ 
\rowcolor{gray!15} 
\hyperlink{det:78}{78} & Uni10 & \yes & \yes & \yes & \yes & \no\\ 
\hyperlink{det:79}{79} & xerus & \no & \no & \yes & \yes & \yes\\ 

    \bottomrule
    \caption{\label{tab:fucntionality-con}Packages that support Contractions (Con).}
\end{longtable}
\newpage
\begin{longtable}{llcccc}
    \toprule
    ID & Name & \multicolumn{4}{c}{Decompositions} \\
    & & CP & Tucker & TensorTrain & Other \\
    \midrule
    \endhead
    \hyperlink{det:1}{1} & AdaTM & \yes & \no & \no & \no\\ 
\rowcolor{gray!15} 
\hyperlink{det:3}{3} & BTAS & \yes & \yes & \no & \no\\ 
\hyperlink{det:7}{7} & CP-CALS & \yes & \no & \no & \no\\ 
\rowcolor{gray!15} 
\hyperlink{det:8}{8} & CSTF & \no & \no & \no & \yes\\ 
\hyperlink{det:12}{12} & D-Tucker & \no & \yes & \no & \no\\ 
\rowcolor{gray!15} 
\hyperlink{det:13}{13} & DFacTo & \yes & \no & \no & \no\\ 
\hyperlink{det:15}{15} & ExaTN & \no & \no & \yes & \no\\ 
\rowcolor{gray!15} 
\hyperlink{det:18}{18} & Genten & \yes & \no & \no & \no\\ 
\hyperlink{det:19}{19} & GigaTensor & \yes & \no & \no & \no\\ 
\rowcolor{gray!15} 
\hyperlink{det:21}{21} & ITensor & \no & \no & \yes & \no\\ 
\hyperlink{det:26}{26} & multiway & \yes & \yes & \no & \yes\\ 
\rowcolor{gray!15} 
\hyperlink{det:27}{27} & N-way toolbox & \yes & \yes & \no & \yes\\ 
\hyperlink{det:32}{32} & ParCube & \yes & \no & \no & \no\\ 
\rowcolor{gray!15} 
\hyperlink{det:33}{33} & ParTensor & \yes & \no & \no & \no\\ 
\hyperlink{det:34}{34} & ParTI! & \yes & \yes & \no & \no\\ 
\rowcolor{gray!15} 
\hyperlink{det:35}{35} & PLANC & \yes & \no & \no & \no\\ 
\hyperlink{det:36}{36} & PLS toolbox & \yes & \yes & \no & \no\\ 
\rowcolor{gray!15} 
\hyperlink{det:37}{37} & Pytensor & \no & \yes & \no & \no\\ 
\hyperlink{det:40}{40} & rTensor & \yes & \yes & \no & \yes\\ 
\rowcolor{gray!15} 
\hyperlink{det:41}{41} & rTensor (randomized) & \yes & \no & \no & \no\\ 
\hyperlink{det:42}{42} & scikit-tensor & \yes & \yes & \no & \yes\\ 
\rowcolor{gray!15} 
\hyperlink{det:43}{43} & Scikit-TT & \no & \no & \yes & \no\\ 
\hyperlink{det:44}{44} & SPALS & \yes & \no & \no & \no\\ 
\rowcolor{gray!15} 
\hyperlink{det:45}{45} & SPARTan & \no & \no & \no & \yes\\ 
\hyperlink{det:46}{46} & SPLATT & \yes & \no & \no & \no\\ 
\rowcolor{gray!15} 
\hyperlink{det:47}{47} & SuSMoST & \no & \no & \yes & \yes\\ 
\hyperlink{det:48}{48} & T3F & \no & \no & \yes & \no\\ 
\rowcolor{gray!15} 
\hyperlink{det:54}{54} & TDALAB & \yes & \no & \no & \no\\ 
\hyperlink{det:55}{55} & TeNPy & \no & \no & \yes & \no\\ 
\rowcolor{gray!15} 
\hyperlink{det:56}{56} & Tensor Fox & \yes & \no & \no & \no\\ 
\hyperlink{det:57}{57} & Tensor package & \yes & \no & \no & \no\\ 
\rowcolor{gray!15} 
\hyperlink{det:58}{58} & Tensor Toolbox & \yes & \yes & \no & \yes\\ 
\hyperlink{det:59}{59} & tensor{\_}decomposition & \yes & \yes & \no & \no\\ 
\rowcolor{gray!15} 
\hyperlink{det:60}{60} & TensorBox & \yes & \yes & \no & \yes\\ 
\hyperlink{det:61}{61} & TensorD & \yes & \yes & \no & \no\\ 
\rowcolor{gray!15} 
\hyperlink{det:63}{63} & TensorLab & \yes & \yes & \no & \yes\\ 
\hyperlink{det:64}{64} & TensorLab+ & \yes & \no & \no & \yes\\ 
\rowcolor{gray!15} 
\hyperlink{det:65}{65} & TensorLy & \yes & \yes & \yes & \yes\\ 
\hyperlink{det:69}{69} & Three-Way & \yes & \yes & \no & \no\\ 
\rowcolor{gray!15} 
\hyperlink{det:72}{72} & TNR & \no & \no & \no & \yes\\ 
\hyperlink{det:74}{74} & TT-Toolbox & \no & \no & \yes & \no\\ 
\rowcolor{gray!15} 
\hyperlink{det:79}{79} & xerus & \no & \no & \yes & \no\\ 

    \bottomrule
    \caption{\label{tab:fucntionality-decomp}Packages that support Decompositions (Decomp).}
\end{longtable}
\newpage

\begin{longtable}{lllccccc}
  \toprule
  ID & Name & Language & \multicolumn{5}{c}{Functionality} \\
  & & & DatM & EWOps & SpecCon & Con & Decomp \\
  \midrule
  \endhead
  \hyperlink{det:2}{2} & Boost.uBlas.Tensor & {\CC} & \yes & \yes & \yes & \yes & \no\\ 
\rowcolor{gray!15} 
\hyperlink{det:3}{3} & BTAS & {\CC} & \yes & \yes & \yes & \yes & \yes\\ 
\hyperlink{det:9}{9} & CuTensor & C, CUDA & \yes & \yes & \yes & \yes & \no\\ 
\rowcolor{gray!15} 
\hyperlink{det:11}{11} & Cyclops & {\CC} & \yes & \yes & \yes & \yes & \no\\ 
\hyperlink{det:14}{14} & Eigen Tensor & {\CC} & \yes & \yes & \yes & \yes & \no\\ 
\rowcolor{gray!15} 
\hyperlink{det:15}{15} & ExaTN & {\CC}, Py\textsuperscript{i} & \yes & \yes & \yes & \yes & \yes\\ 
\hyperlink{det:16}{16} & Fastor & {\CC} & \yes & \yes & \yes & \yes & \no\\ 
\rowcolor{gray!15} 
\hyperlink{det:17}{17} & FTensor & {\CC} & \yes & \yes & \yes & \yes & \no\\ 
\hyperlink{det:21}{21} & ITensor & {\CC}, Julia & \no & \yes & \yes & \yes & \yes\\ 
\rowcolor{gray!15} 
\hyperlink{det:24}{24} & MATLAB & Matlab & \yes & \yes & \yes & \yes & \no\\ 
\hyperlink{det:30}{30} & NumPy & Python & \yes & \yes & \yes & \yes & \no\\ 
\rowcolor{gray!15} 
\hyperlink{det:34}{34} & ParTI! & C, CUDA, Mat\textsuperscript{x} & \yes & \yes & \yes & \no & \yes\\ 
\hyperlink{det:37}{37} & Pytensor & Python & \yes & \yes & \yes & \yes & \yes\\ 
\rowcolor{gray!15} 
\hyperlink{det:38}{38} & PyTorch & Python, {\CC}, CUDA & \yes & \yes & \yes & \yes & \no\\ 
\hyperlink{det:40}{40} & rTensor & R & \yes & \yes & \yes & \no & \yes\\ 
\rowcolor{gray!15} 
\hyperlink{det:42}{42} & scikit-tensor & Python & \yes & \yes & \yes & \no & \yes\\ 
\hyperlink{det:49}{49} & TACO & {\CC}, {\CC} {\textrightarrow} {\CC} & \yes & \yes & \yes & \yes & \no\\ 
\rowcolor{gray!15} 
\hyperlink{det:50}{50} & TAL{\_}SH & C, {\CC}, Fort & \yes & \yes & \yes & \yes & \no\\ 
\hyperlink{det:51}{51} & TBlis & {\CC} & \yes & \yes & \yes & \yes & \no\\ 
\rowcolor{gray!15} 
\hyperlink{det:58}{58} & Tensor Toolbox & Matlab & \yes & \yes & \yes & \yes & \yes\\ 
\hyperlink{det:62}{62} & TensorFlow & {\CC}, Python & \yes & \yes & \yes & \yes & \no\\ 
\rowcolor{gray!15} 
\hyperlink{det:65}{65} & TensorLy & Python & \yes & \yes & \yes & \yes & \yes\\ 
\hyperlink{det:67}{67} & TensorOperations.jl & Julia & \yes & \yes & \yes & \yes & \no\\ 
\rowcolor{gray!15} 
\hyperlink{det:70}{70} & TiledArray & {\CC} & \yes & \yes & \yes & \yes & \no\\ 
\hyperlink{det:78}{78} & Uni10 & {\CC} & \yes & \yes & \yes & \yes & \no\\ 

  \bottomrule
  \caption{\label{tab:fucntionality-complete}Packages that offer functionality in  at least four out of the
    five categories listed in Section~\ref{sec:table}.}
\end{longtable}

%\newpage
%\begin{longtable}{lllccccc}
%	\toprule
%	ID & Name & Language & \multicolumn{5}{c}{Functionality} \\
%	& & & DatM & EWOps & SpecCon & Con & Decomp \\
%	\midrule
%	\endhead
%	\input{data/table-func-spec.data}
%	\bottomrule
%	\caption{\label{tab:fucntionality-spec}Packages that specialize in a particular type of general tensor operations, such as transposition or contraction.}
%\end{longtable}

\subsection{Notable omissions}
Certain packages that are well known in the community for offering tensor contractions and other operations include TAMM~\cite{TAMM} and TCE \cite{TCE}.
These packages were not included in the list, since both are implemented as components of a larger project, NWChem
\cite{NWChem} (a software primarily targeted towards computational chemistry), and are not usable indepentently.
Furthermore, while  many tensor operations can be cast in terms of BLAS and LAPACK calls, e.g.,~\cite{Di_Napoli2014:210,Di_Napoli2017:318,Peise2015:380}, in this survey we only focus on packages that support multi-dimensional arrays.

%%% Local Variables:
%%% mode: latex
%%% TeX-master: "../Tensor_SW"
%%% End:

\section{Conclusion}\label{sec:conclusion}

We provide a survey of tensor packages which arise in a wide range of fields and applications.
Most of the packages are written in (or target) one of a handful of well-known programming languages, and are standalone, i.e.,
they do not depend on one another. This means that 
% Some of the packages have dependencies among one another, but the majority are standalone and
many packages (re-)implement, often sub-optimally, the same or similar functionality within their own codebase
(e.g., tensor transposition, and specific operations such as MTTKRP and TTV).

With this list we aim to help both (new) users in finding a suitable package for their needs, and developers in
identifying opportunities for cooperation, modularity, and optimization.
Ultimately, our goal is to create awareness about the level of redundancy that permeates the software landscape of
tensor computations, and the potential implications on software quality, performance, and productivity.

Furthermore, we see this survey as a first step to pave the way towards a set of universal, optimized, building blocks,
which shall play the same role as the one that BLAS and LAPACK have played (and are playing) in the domain of numerical linear
algebra. 

%%% Local Variables:
%%% mode: latex
%%% TeX-master: "../Tensor_SW"
%%% End:

%\appendix
%\section{An example appendix}
%\lipsum[71]
%
%\begin{lemma}
%Test Lemma.
%\end{lemma}

\section*{Acknowledgments}
Financial support from the Deutsche Forschungsgemeinschaft (German Research Foundation) through grant IRTG 2379 is gratefully acknowledged.

\appendix
\section{Questionnaire} \label{app:questionnaire}

Any cross-disciplinary investigation of this size is bound to be incomplete and to contain mistakes. 
We therefore kindly ask the reader to help us by emailing corrections and additions to \email{pauldj@cs.umu.se}.
When providing information about a package, please consider the following questions.
\begin{enumerate}
\item What is the name of the package?
\item Where is the source code located?
\item Would you please provide a reference (preferably in BibTeX format) to a publication, preprint, or website about
  the package?
\item In which programming language(s) is the package written?
\item Which programming language(s) do users have to code in to use the package?
\item Is the package standalone? Alternatively, does it depend on another package that is either in the list or that belongs in the list?
\item What is the target computing architecture (CPUs, GPUs, Distributed Memory, others)?
\item Does the package support layout-related operations, such as tensor transpositions, or reshaping, \dots?
\item Does it support element-wise operations, such as addition/subtraction, reductions, \dots?
\item Does it support general binary contractions? If not, what are the limitations?
\item Does it support only a specific, subset of contractions? (e.g., TTV, TTM, MTTKRP)
\item Does it support other types of contractions? Which one(s)?
\item Does it support tensor decompositions? If so, please provide a comma separated list of decompositions supported.
\item Would you please describe in 1-2 sentences what the package is about, what target problem it addresses, and what functionalities it provides?
\item Is there any other information about the package that you deem essential in order to describe its functionality?
\end{enumerate}
Thank you for your help!

%%% Local Variables:
%%% mode: latex
%%% TeX-master: "../Tensor_SW"
%%% End:

\bibliographystyle{lib/siamplain}
\bibliography{TensorSW}

\end{document}